\documentclass[preprint]{aastex}
\usepackage{graphicx,amssymb,amsmath,times,verbatim}
\usepackage{subfigure}
\usepackage{color}

\usepackage{amssymb,amsmath}
\usepackage[normalem]{ulem}

\newcommand{\source}{PKS 1510-089}

\def\lta{\mathrel{\spose{\lower 3pt\hbox{$\mathchar"218$}}
     \raise 2.0pt\hbox{$\mathchar"13C$}}}
\def\gta{\mathrel{\spose{\lower 3pt\hbox{$\mathchar"218$}}
     \raise 2.0pt\hbox{$\mathchar"13E$}}}

\def\mathnew{\mathsurround=0pt}

\def\simov#1#2{\lower .5pt\vbox{\baselineskip0pt \lineskip-.5pt
\ialign{$\mathnew#1\hfil##\hfil$\crcr#2\crcr\sim\crcr}}}

\shorttitle{Flux Variability in FSRQ PKS 1510-089}
\shortauthors{Kushwaha et al.}

\begin{document}


\title{Evidence for two Lognormal States in Multi-wavelength Flux Variation of FSRQ PKS 1510-089}
\author{ Pankaj Kushwaha\altaffilmark{1}, Sunil Chandra\altaffilmark{2}, Ranjeev Misra\altaffilmark{1},
S. Sahayanathan\altaffilmark{3}, K. P. Singh\altaffilmark{2}, K. S. Baliyan\altaffilmark{4}}

\altaffiltext{1}{Inter University Center for Astronomy and Astrophysics, Pune 411007, India;
pankajk@iucaa.in}
\altaffiltext{2}{Department of Astronomy \& Astrophysics, Tata Institute of Fundamental
Research, Mumbai 400005, India}
\altaffiltext{3}{Astrophysical Sciences Division, Bhabha Atomic Research Centre, Mumbai
400085, India}
\altaffiltext{4}{Physical Research Laboratory, Ahemdabad 380009, India}

\begin{abstract}
We present a systematic characterization of multi-wavelength emission from blazar
PKS 1510-089 using well-sampled data at infrared(IR)-optical, X-ray and $\gamma$-ray
energies. The resulting flux distributions, except at X-rays, show two distinct
lognormal profiles corresponding to a high and a low flux level. The dispersions
exhibit energy dependent behavior except for the LAT $\gamma$-ray and optical B-band.
During the low level flux states, it is higher towards the peak of the spectral energy distribution,
with $\gamma$-ray being intrinsically more variable followed by IR and then optical,
consistent with mainly being a result of varying bulk Lorentz factor. On the other hand,
the dispersions during the high state are similar in all bands expect optical B-band, 
where thermal emission still dominates. The centers of distributions are  a factor of
$\sim 4$ apart, consistent with anticipation from studies of extragalactic
$\gamma$-ray background with the high state showing a relatively harder mean spectral
index compared to the low state.
\end{abstract}


\section{INTRODUCTION} \label{sec: intro}
Flux variability across the electromagnetic spectrum over a wide range of timescales
ranging from fractions of second to years is an important characteristic of astrophysical
objects powered by active black holes. The temporal variability has been found to be
stochastic and widely believed to be a result of a complex and dynamic interplay
between the magnetic field, black hole spin, and accreting plasma as well as
relativistic jet (if present). In fact, well sampled time series analysis of
these sources show similar features albeit at different timescales, lending support
to understanding that similar processes are powering the emission at the base level
\citep[eg.][]{2002PASJ...54L..69N, 2003ICRC....5.2729A}. Thus, based on these and other
similarities, active galactic nuclei (AGN), hosting supermassive black holes \citep[SMBHs,][]{2004MNRAS.348..783M}
have been claimed to be a scaled-up versions of Galactic X-ray binaries hosting
stellar mass black holes \citep{2006Natur.444..730M}. 

Blazars, consisting of BL Lacertae objects (BLL) and flat spectrum radio quasars (FSRQs)
are subclass of radio-loud AGN with emission dominated by relativistic
jet beamed towards us. They are characterized by a high flux variability across the
entire accessible electromagnetic spectrum on timescales ranging from minutes to
years with a high degree of polarization at radio and optical energies which often
changes with changes in the source flux states. Additionally, high-resolution imaging
in radio wavebands show superluminal features across the jet \citep{2013AJ....146..120L}.
Despite a wide range in temporal variability timescales, their spectral energy distributions
(SEDs) show a characteristic broad double-hump profile with a low energy peak
between infrared (IR) to ultraviolet (UV)/X-rays while the other at $\gamma$-ray energies.
Emission from other components e.g. disk, corona, torus, line emitting regions etc. have
also been observed in few sources but almost always sub-dominant compared to the
jet emission. The low energy hump of the SED is well understood
to be synchrotron radiation from relativistic non-thermal distribution of electrons
(or pairs), while two schools of thought: leptonic and hadronic, exist for emission at 
the high energy hump. The former attributes the high energy emission to inverse Compton
processes \citep{2014Natur.515..376G, 2015MNRAS.453.4070P} while the latter attributes
it to hadron-photon initiated cascades and/or proton synchrotron
\citep{1998Sci...279..684M, 2003APh....18..593M}. Irrespective of the origin, emission from
blazars have been found to be stochastic, similar to other AGN and Galactic X-ray 
binaries \citep{2008bves.confE..14M, 2010ApJ...722..520A, 2012ApJ...749..191C,
2013ApJ...773..177N, 2014ApJ...786..143S} with reports of quasi-periodicity as well
\citep{2008Natur.455..369G}. Study of long-term
behavior of variability in blazars, however, is highly dependent on the simultaneous
coverage of source over a broad spectrum. But, for $\gamma$-ray bright blazars, the
continuous scanning of entire sky every $\sim$ 3 hours between 0.1 - 300 GeV by
the \emph{Fermi}-LAT and coordinated multi-wavelength follow-ups by observatories
supporting \emph{Fermi} provides an unprecedented amount of data for such studies.

Flares are loosely defined as the brightening of source emission regardless of the
nature of physical processes causing the brightening, and was first introduced
in the context of emission from the Sun \citep[e.g.,][]{2008LRSP....5....1B}. It has a
different meaning in context of different class of sources and spans a wide range in
spectral/flux and temporal domains. In the spectral/flux domain, the distribution
of various flares characteristics varies dramatically across different sources e.g. 
Solar flare peak X-ray fluxes show a power-law distribution (\citet
{2012ApJ...754..112A}, see also \citet{2003ICRC....5.2729A}) while Galactic X-ray
binaries exhibit a lognormal distribution on sub second timescales
\citep[e.g.][]{2005MNRAS.359..345U}. In blazars, flares are generally defined using the flux doubling and
halving timescales \citep{2013MNRAS.430.1324N}, and encode the characteristics of
processes happening in the jet. As a result, flux distribution of blazars in different
activity states has been investigated in many studies to understand the nature of
emission processes \citep{2009A&A...503..797G, 2010A&A...520A..83H, 2015arXiv150903104C}
and jet connection with central engine. However, a systematic study of flux
distribution has been lacking in blazars, mainly due to inhomogeneous sampling and
their stochastic variability nature. Additionally, such studies demand a well 
sampled source variability states including both flaring and quiescent states
to perform a meaningful statistical analysis \citep[eg.][]{2010A&A...520A..83H}.

Here, we report a systematic study of the long-term multi-wavelength flux 
distribution of FSRQ \source,~located at the redshift of $\rm 0.36$. The study is
made feasible due to near-continuous detection of \source~in the LAT band over a
3-days time-bin. Further, it shows many flaring states compared to other
sources with  similar continuous detections in LAT (in preparation). This, in turn, also allowed
us to access the sampling effects in different energy bands and thus,
validate the robustness of our results.

\section{Observations and Data Reduction}
PKS 1510-089 is one of the  most extensively monitored blazars at all wavelengths
accessible from the ground and space under the \emph{Fermi AGN Science:
Multiwavelength Observing - Support
Programs}\footnote{http://fermi.gsfc.nasa.gov/ssc/observations/multi/programs.html}
to understand the enigmatic AGN. The continuous scanning of the sky by \emph{Fermi}
and the corresponding multi-wavelength near-continuous follow-ups from the ground 
have generated vast amount of well sampled data. Since blazar variability
is stochastic \citep{2008bves.confE..14M, 2012ApJ...749..191C, 2013ApJ...773..177N,
2014ApJ...786..143S}, we have mainly
used simultaneous/contemporaneous multi-wavelength data from $\gamma$-ray to optical-IR
except at X-rays where we have used RXTE due to lack of well sampled data by the 
\emph{Swift} observatory. The details of data archives and resources used are summarized
in Table \ref{tab:ObsLog}.

\begin{table}[!ht]
 \caption{Multi-wavelength data and resources.}
 \begin{tabular}{ccccc}
 \hline \hline
 Facility & Energy Range & Data time period & No. of data points & Data* \\
 \hline
Fermi        & {\bf 0.1 - 300 GeV} & 2008-08-05 to 2015-09-10 & 840 (3-days bin) & 840 (97.2\%) \\
RXTE         & 3.0 - 20.0 keV  &  1996-12-13 to 2011-12-30 & 1316 & 1137 (62.1\%) \\
XRT (Swift)        & 0.3-10.0 keV & 2006-08-05 to 2013-09-24 & 238 & - \\
SMARTS & B & 2008-05-17 to 2015-07-01 & 592 & 385 (44.4 \%) \\
       & V & 2008-07-15 to 2015-07-01 & 587 & 381 (44.9 \%) \\
       & R & 2008-02-05 to 2015-07-01 & 592 & 386 (42.8 \%) \\
       & J & 2008-02-05 to 2015-07-01 & 576 & 376 (41.7 \%) \\
       & K & 2009-04-05 to 2015-07-01 & 517 & 341 (44.9 \%) \\
\hline
 \end{tabular}
 \\
 {$^*$ Number of data points after binning uniformly over 3-days for
 the mentioned duration and its percentage in bracket.}
\label{tab:ObsLog}              
\end{table}

\subsection{LAT $\gamma$-ray data}
The \textit{PASS8} Fermi LAT $\gamma$-ray data for the duration of MJD 54683 -
57275 (Table \ref{tab:ObsLog}) is analyzed using \emph{Fermi}-LAT Science tool
version v10r0p5 following the recommended analysis procedures\footnote{http://fermi.\
gsfc.nasa.gov/ssc/data/analysis/scitools/python\_tutorial.html}. Events, more likely
to be photons, classified as ``evclass=128, evtype=3'', energy $>$ 100 MeV and
zenith angles $<$ 90$^\circ$ were selected from a region of interest (ROI) of $15^\circ$
centered at the source. The corresponding good time intervals were calculated
using the recommended selection ``(DATA\_QUAL$>$0)\&\&(LAT\_CONFIG==1)''. Effects of
selections on data and sources outside the ROI were accounted by generating exposure
map on ROI, and an additional annulus of $10^\circ$ around ROI. These events were
then analyzed using \emph{unbinned maximum likelihood} method (PYTHON implementation
of \emph{gtlike}) with \textit{PASS8} instrument response function (\emph{P8R2\_SOURCE\_V6})
and input source model from the 3rd LAT catalog \citep[3FGL -- gll\_psc\_v16.fit;][]
{2015ApJS..218...23A}. Contributions of Galactic and isotropic emission were accounted
by using the latest Galactic diffuse emission (gll\_iem\_v06.fits) and isotropic
background (iso\_P8R2\_SOURCE\_V6\_v06.txt) templates provided by the LAT science team.

The three day averaged light curve in 0.1--300 GeV is generated following the above
procedures assuming a log-parabola model [$dN/dE = N_0 (E/E_b)^{-\alpha-\beta log(E/E_b)}$] 
with normalization ($\rm N_0$) and spectral indices ($\rm \alpha, ~\beta$) free to vary.
The likelihood fit was performed iteratively by removing point sources with Test
Statistic (TS) $<$ 0 until it converged as done in \citet{2014ApJ...796...61K}. However,
for the scientific analysis, only fluxes corresponding to TS $>$ 9 ($\approx$ 3
$\sigma$) have been used, resulting in a near-continuous detection of the source
except for 3\% of the total duration.

\subsection{RXTE quick look data}
The 1316 spectra derived from quick look RXTE data are individually modeled in
\textit{XSPEC} using $\chi^2$ statistics with a minimum of 20 counts per bin.
The \textit{wabs*power-law} model with a fixed $\rm N_H$ \citep[7.4 $\times$ 10$^{20}$
cm$^{-2}$;][]{2015A&A...578A..78K} was used to fit and estimate the 3.0-20.0 keV unabsorbed fluxes.

\subsection{SMARTS long-term IR-optical monitoring data}
\source~ is among the blazars that are regularly monitored at almost
on daily cadence by the group at Yale University under the banner of \emph{SMARTS}
consortium. The reduced \emph{SMARTS} data for all five optical/NIR bands (BVRJK)
are publicly available and the details of reduction methodology
are discussed in \citet{2012ApJ...756...13B}. Here, we have used 
data from all the bands after correcting for Galactic reddening \citep{2011ApJ...737..103S} 
with an E(B-V) of 0.085 $\pm$ 0.004.

\section{ANALYSIS AND RESULTS}
Multi-wavelength flux distribution of sources with broadband stochastic variability
is a unique tool to understand and probe the nature of physical processes eg. a
normal flux distribution suggests additive processes while a lognormal refers to
cascade/multiplicative processes. Additionally, it offers valuable insights in
unifying the processes if activities in different bands are correlated, but 
requires a well sampled multi-wavelength data. In 
\source, the $\gamma$-ray activity is well sampled and almost every $\gamma$-ray
activity\footnote{Visual inspection of multi-wavelength light curves only}. is
reflected in the IR-optical bands\footnote{Few
non-associated events will have no bearing on general trends.}. Further, despite gaps
in the observations, the IR-Optical data are well sampled in terms of flux states of
the source. X-ray/UV data by \emph{Swift}, on the other hand, are relatively few
and belong mostly to the elevated activity state. Hence, we have
used the non-contemporaneous but well sampled \emph{RXTE} data at X-rays.

The multi-wavelength flux characterization is performed by constructing
histograms (normalized) of logarithm\footnote{unless stated otherwise, logarithm
in the present work refers to logarithm of physical quantity to the base 10.} of flux
in different energy bands from IR to $\gamma$-rays. An equi-spaced flux histogram
of 20  bins has been found to be satisfactory for all the bands\footnote{Also consistent
with the available statistical methods for generating histograms in literature, like
Knuth, Bayesian etc.}.
Additionally, since multi-wavelength fluxes appear correlated, we have also
constructed the intrinsic variability and SED associated with the resulting flux
distribution to better understand the underlying processes.

\subsection{Flux Distribution: Log-normality} \label{subsec:flxHist}

\subsubsection{Gamma-ray Flux Histogram: LAT}\label{subsub:LAT}
\begin{figure}[!h]
\begin{minipage}[b]{0.46\linewidth}
\centering
\subfigure[LAT]{\includegraphics[scale=0.7]{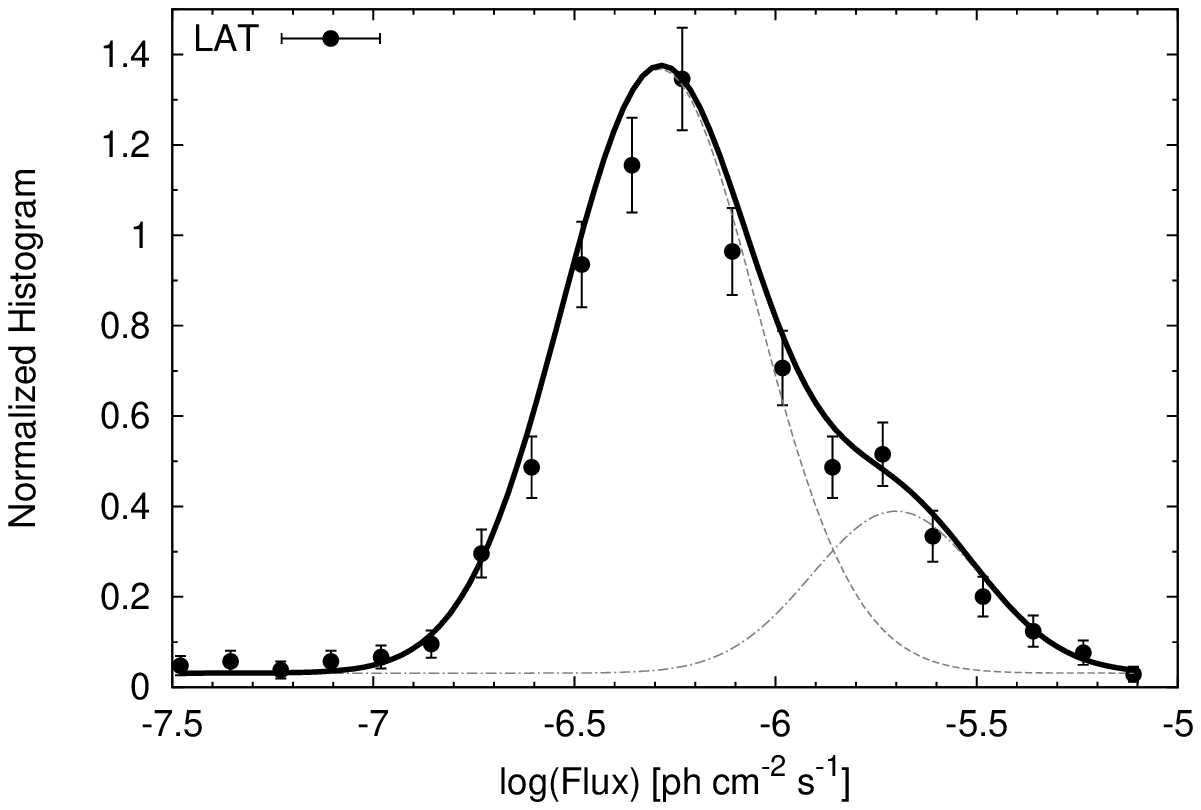}}
\hfill
\subfigure[RXTE]{\includegraphics[scale=0.7]{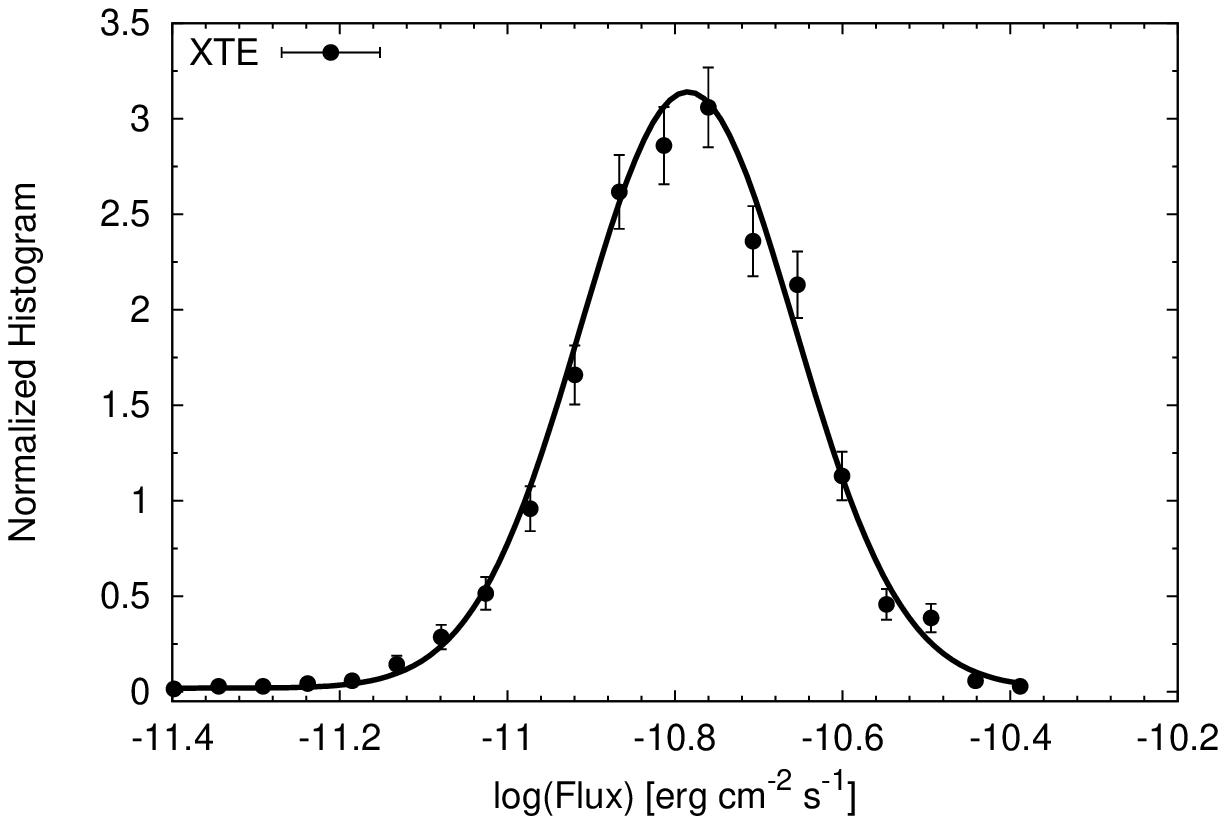}}
\end{minipage}
\quad
 \begin{minipage}[b]{0.50\linewidth}
 \centering
\subfigure[SMARTS]{\includegraphics[scale=0.67]{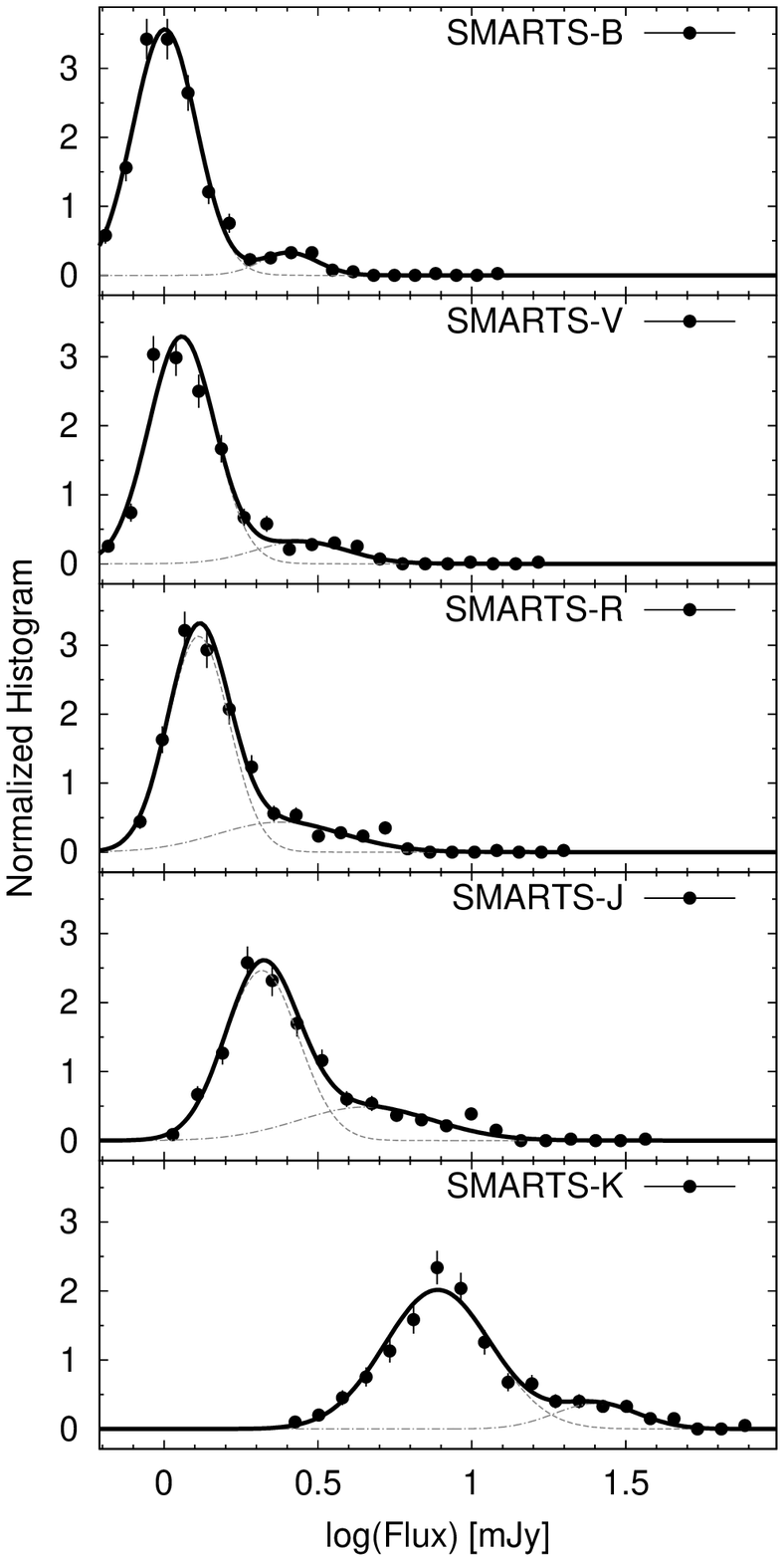}}
\end{minipage}
\caption{Multi-wavelength flux histograms of \source~ from IR to $\gamma$-ray
energies (see \S\ref{subsec:flxHist}).}
\label{fig:LAT2XHist}
\end{figure}

The histogram (normalized) of the logarithmic photon flux from LAT is shown in the
top-left panel of Figure \ref{fig:LAT2XHist}. The profile of the resulting histogram is
consistent with two Normal distributions of general form \citep[e.g.][]{2010A&A...520A..83H}
\begin{equation}
 \rm f(x) = \frac{a}{\sqrt{2\pi \sigma_0^2}} e^\frac{-(x-x_0)^2}{2\sigma_0^2} \
       + \frac{(1-a)}{\sqrt{2\pi \sigma_1^2}} e^\frac{-(x-x_1)^2}{2\sigma_1^2} +c
\end{equation}
where $\rm a$ is the normalization, $\rm x_0$ and $\rm x_1$ are the centers of
the distribution with widths $\rm \sigma_0$ and $\rm \sigma_1$, respectively. $c$ is
a constant offset to account for the base flux and/or long tail, if any,
and has been fixed to zero in case of best fit value being negative or consistent
with zero. The best fit values of parameters along with 1$\sigma$
uncertainties are given in Table \ref{tab:2GaussPar}.

\subsubsection{X-ray Flux Histogram: RXTE}\label{subsub:XTE}
Similarly, the histogram of the X-ray flux from RXTE
is shown in the bottom-left panel of Figure \ref{fig:LAT2XHist}. Most of the data are
non-contemporaneous to the $\gamma$-ray observations and contrary to $\gamma$-ray flux
distribution, the X-ray histogram is consistent with a single lognormal distribution (see 
Table \ref{tab:2GaussPar}).

\subsubsection{IR-Optical Flux Histogram: SMARTS}\label{subsub:SMARTS}
The flux distributions from all the SMARTS bands are shown in the right panel of
Figure \ref{fig:LAT2XHist}. As is the case with LAT $\gamma$-ray flux distribution,
the SMARTS histograms are also consistent with a bi-lognormal distribution. The corresponding best
fit parameters value are reported in the Table \ref{tab:2GaussPar}.  The relatively
high reduced-$\rm \chi^2$ for V-band appears to be a result of competitive contributions from 
the jet and accretion disk, where the latter hardly varies in comparison to the former,
but contributes significantly except in the very high flux states (tail of
the distribution).

\begin{table}[!ht]
\centering
\caption{Best fit parameter values with 1$\sigma$ errors.}
 \begin{tabular}{lccccccc}
\hline\hline
Energy Band &	a	& $\rm x_0^\ast$ & $\sigma_0^\ast$ &  $\rm x_1^\ast$ & $\sigma_1^\ast$ & dof & $\rm \chi^2/dof$ \\
\hline
LAT 	& 0.81 	$\pm$ 0.06 & -6.28 $\pm$ 0.03 & 0.24 $\pm$ 0.02 & -5.70 $\pm$ 0.08 & 0.21 $\pm$ 0.04 & 14 & 1.17 \\
RXTE 	& 1.0 	$\pm$ ---- & -10.785 $\pm$ 0.003 & 0.128 $\pm$ 0.003 & ---- $\pm$ ---- & ---- $\pm$ ---- & 17 & 1.10 \\
SMARTS-B 	& 0.92 $\pm$ 0.01 & 0.003 $\pm$ 0.004 & 0.103 $\pm$ 0.004 & 0.40 $\pm$ 0.02 & 0.09 $\pm$ 0.02 & 15 & 0.76 \\
SMARTS-V 	& 0.88 $\pm$ 0.04 & 0.057 $\pm$ 0.009 & 0.107 $\pm$ 0.007 & 0.44 $\pm$ 0.07 & 0.14 $\pm$ 0.04 & 15 & 1.88 \\
SMARTS-R 	& 0.78 $\pm$ 0.14 & 0.112 $\pm$ 0.008 & 0.10 $\pm$ 0.01 & 0.38 $\pm$ 0.16 & 0.20 $\pm$ 0.07 & 15 & 1.27 \\
SMARTS-J	& 0.73 $\pm$ 0.12 & 0.32 $\pm$ 0.02 & 0.12 $\pm$ 0.01 & 0.66 $\pm$ 0.12 & 0.22 $\pm$ 0.06 & 15 & 1.40 \\
SMARTS-K 	& 0.88 $\pm$ 0.03 & 0.89 $\pm$ 0.01 & 0.17 $\pm$ 0.01 & 1.40 $\pm$ 0.04 & 0.13 $\pm$ 0.03 & 15 & 1.04 \\
\hline
\multicolumn{6}{l}{$^\ast$ LAT: ph cm$^{-2}$ s$^{-1}$; RXTE: erg cm$^{-2}$ s$^{-1}$; SMARTS: mJy}\\
 \end{tabular}
 \label{tab:2GaussPar}
\end{table}

\subsubsection{Inhomogeneous Sampling and Bias on Histograms}\label{subsec:Bias}

\begin{figure}[!ht]
\centering
\begin{minipage}[t]{0.48\linewidth}
 \includegraphics[scale=0.65]{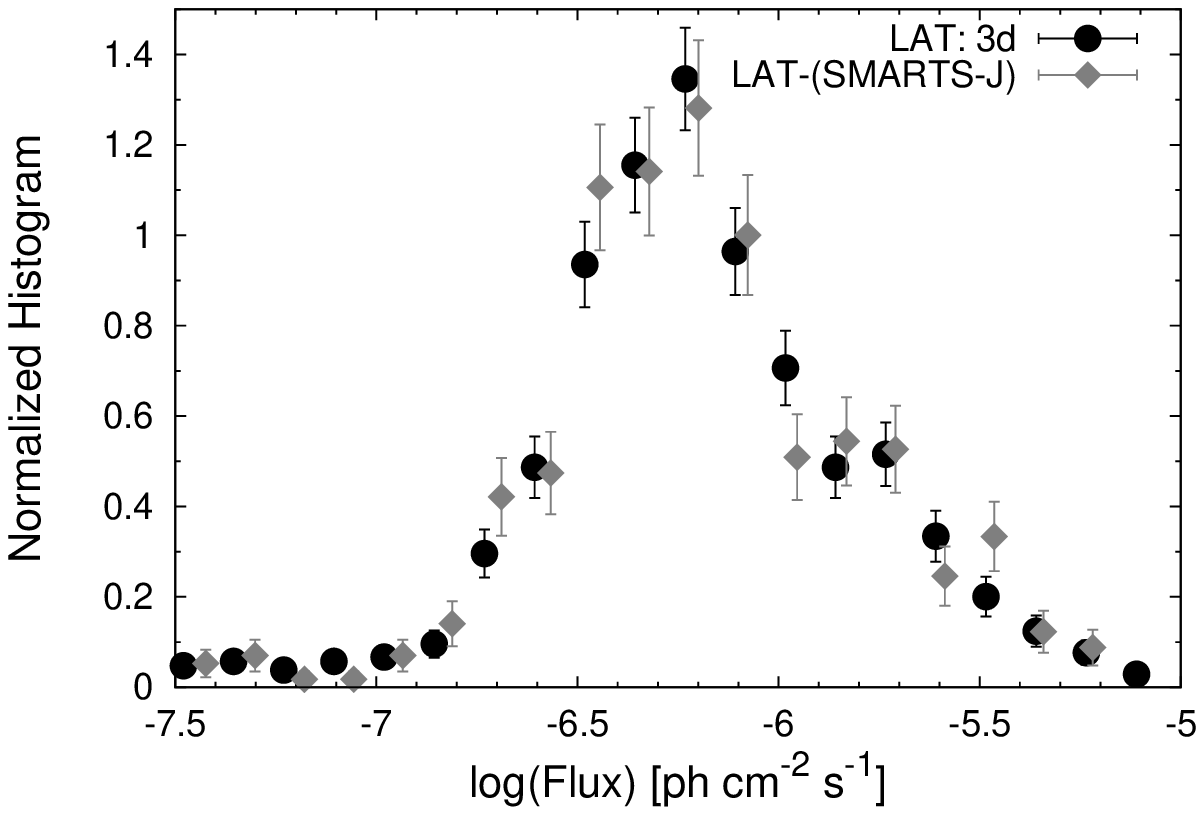}
 \includegraphics[scale=0.65]{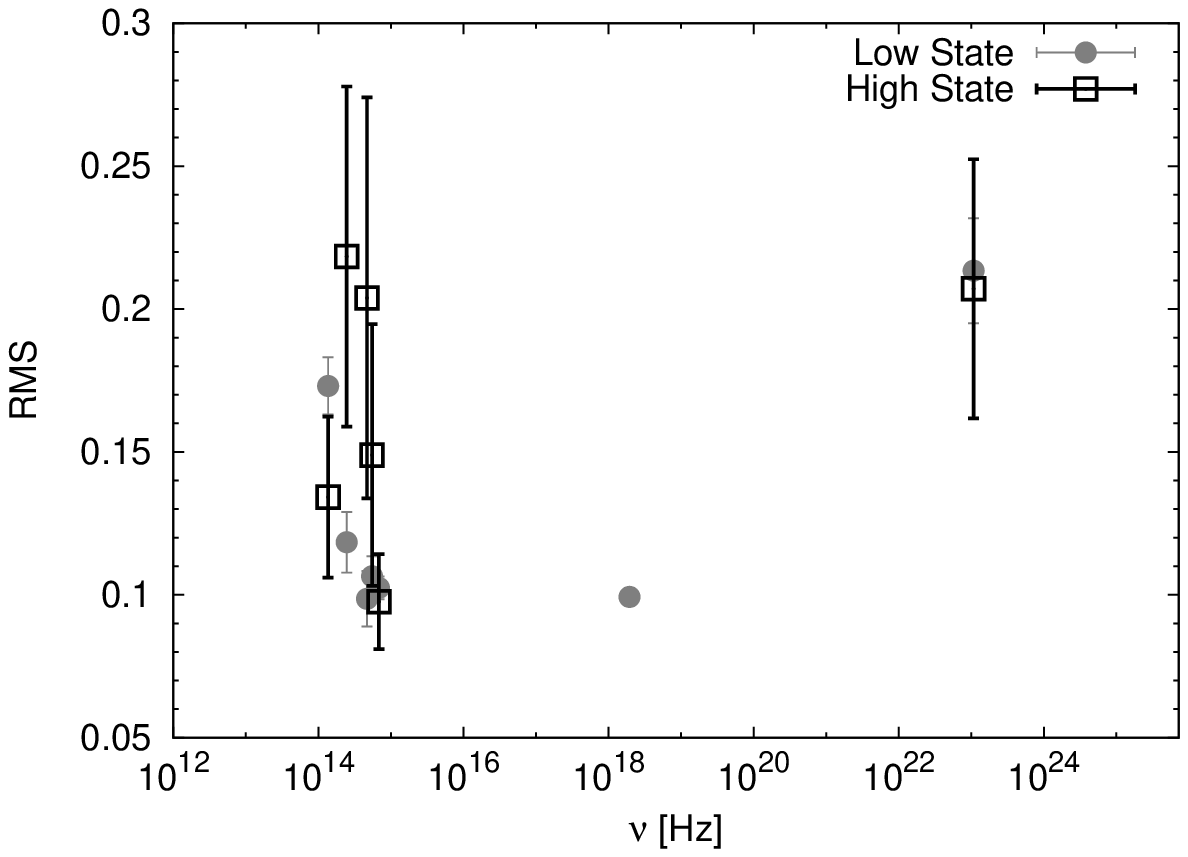}
\end{minipage}
\quad
 \begin{minipage}[t]{0.48\linewidth}
 \includegraphics[scale=0.65]{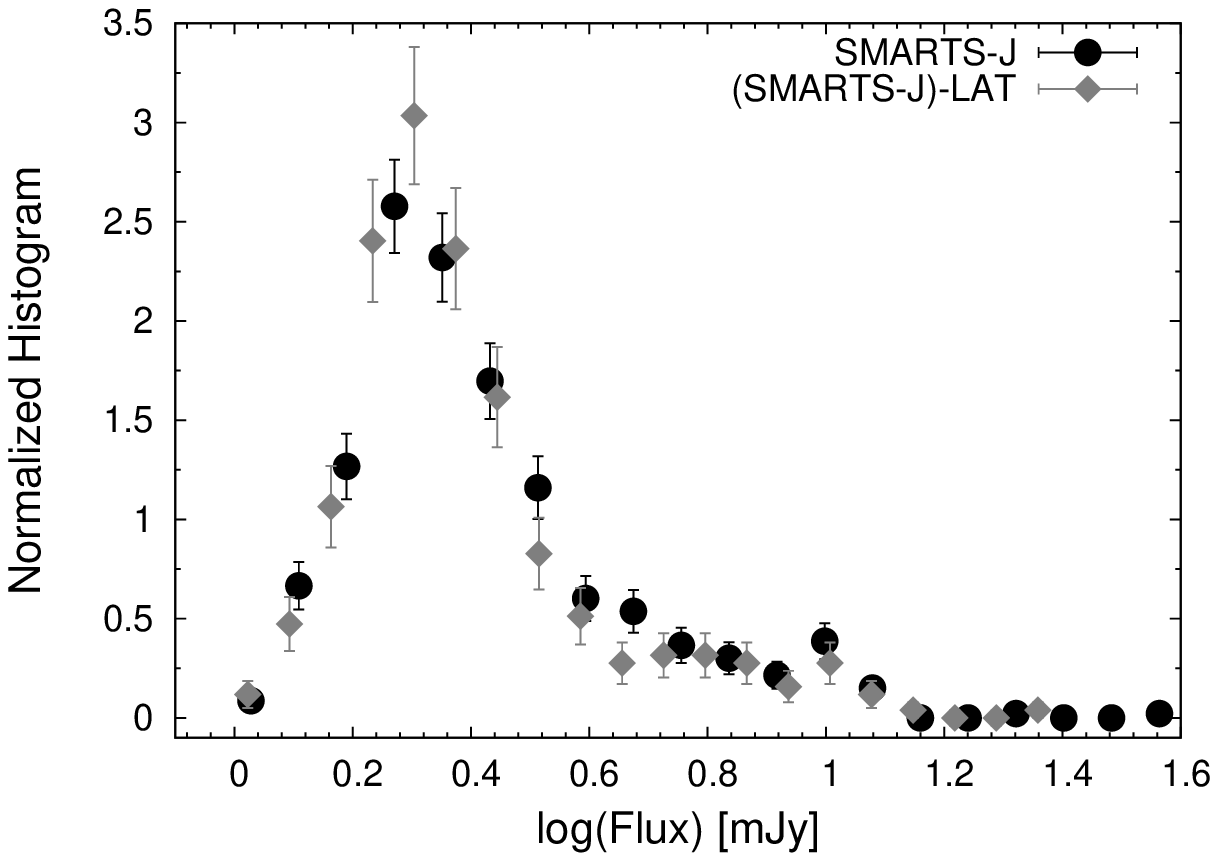}
 \includegraphics[scale=0.65]{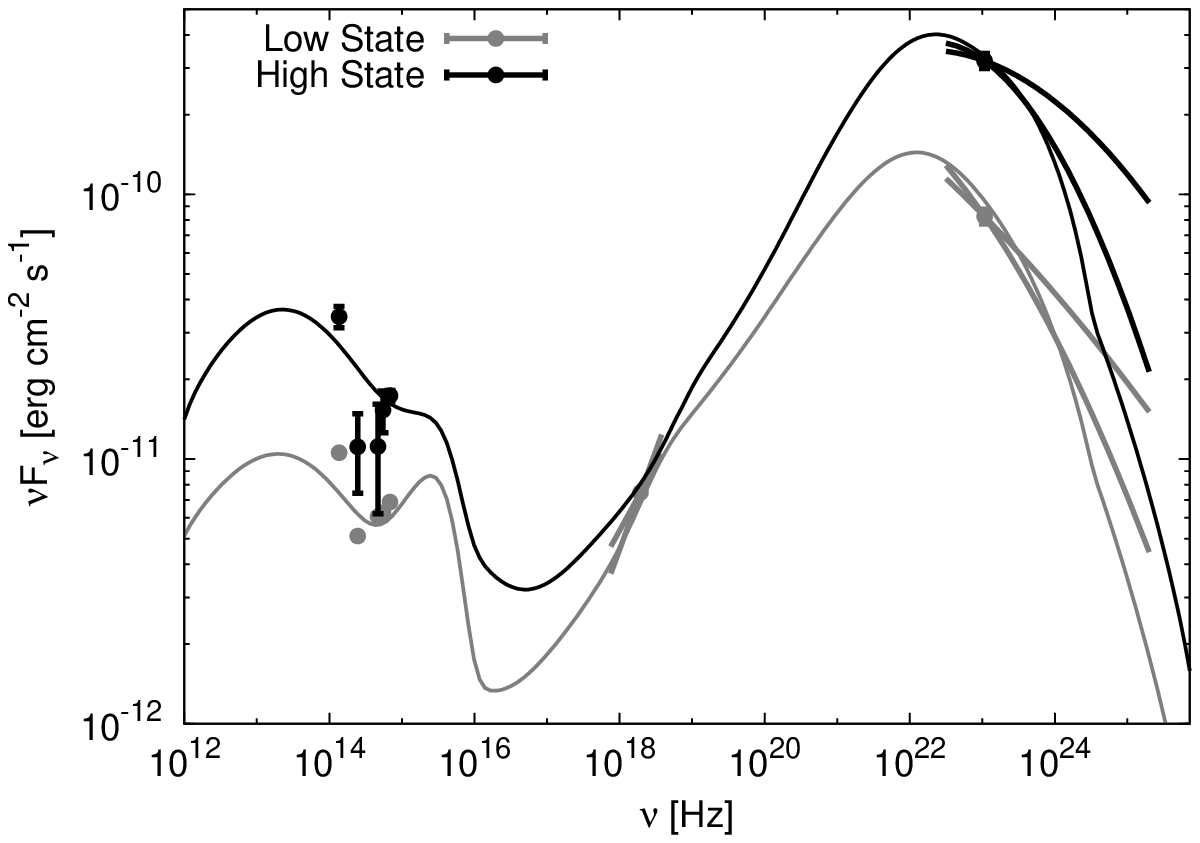}
\end{minipage}
 \caption{\emph{Top}: LAT histograms from original data (left, black) and
 from LAT data sampled at the \emph{SMARTS}-J band durations (left, grey). Similarly,
 \emph{SMARTS}-J histograms from original data (right, black) and from J-band data
 averaged over LAT binning duration of 3-days (right, grey; see \S\ref{subsec:Bias}).
 \emph{Bottom}: The RMS spectra (left) and SEDs (right) corresponding the Low State
 and the High State (see \S\ref{subsec:rmsSED}). SED
 is estimated assuming a $\rm \Lambda CDM$ cosmology with $\Omega_\Lambda=0.7$,
 $\Omega_M=0.3$, and $\rm H_0=70~ km/s/Mpc$.}
 \label{fig:histComSpec}
\end{figure}

The incoherent and in-homogeneous sampling in different energy bands can introduce
bias in the flux distributions, especially at IR-Optical energies where blazars
 also show strong intra-night variability \citep[e.g.][]{2011ApJ...731..118C, 
2007MNRAS.374..357C}. However, for correlated variations, a good estimate of bias
introduced by the sampling can be accessed by using sampling criteria similar to the
best (LAT) and least (SMARTS-J) monitored light curves. For the effect of
inhomogeneous sampling, we have constructed a LAT flux histogram by considering
only those fluxes which have at least one observation in the SMARTS-J band.
Other way round, we have binned SMARTS-J data using LAT timebins and
constructed its histogram. The rebinned histograms along with the original
ones are shown in the top panel of Figure \ref{fig:histComSpec}. Similarly,
we have checked the other \emph{SMARTS} bands and the RXTE data by considering only
the time bins having LAT observations. The best fit parameters to the rebinned
histograms are compatible with the value given in the Table \ref{tab:2GaussPar}
except for V-band which is due to contribution by jet and disk (individually
jet/disk dominated emission show bimodality). 
Thus, our results are independent of sampling effects and hence, are robust.

\subsection{RMS Variability and SED} \label{subsec:rmsSED}
To better understand the flux states, we have also constructed the intrinsic
variability (excess rms) spectra and SEDs corresponding to the two flux states labeled
as Low State and High State using intersection of the two lognormals as the
dividing point. The respective plots are shown in the bottom panels 
of Figure \ref{fig:histComSpec}. The RMS variability spectrum of respective State
is calculated by subtracting the effects of flux errors from the respective best
fit $\sigma$ \citep[$\rm \sigma_0$ and $\rm \sigma_1$;][]{2003MNRAS.345.1271V},
while the SEDs are constructed using the best fit central flux {values, $\rm x_0$
and $\rm x_1$}  and associated 1$\sigma$ uncertainties. A one zone leptonic
model of same emission region size, incorporating synchrotron and IC scattering of
synchrotron, BLR and torus \citep{2008ApJ...672..787K} photon fields
\citep{2013MNRAS.433.2380K, 2012MNRAS.419.1660S} successfully reproduces the SEDs with parameters
values reported in Table \ref{tab:SEDparVal}. In both RMS spectra and SEDs, the
X-ray ({$\Gamma$\footnote{$\rm N(E) = N_0 \left(E/E_0\right)^{-\Gamma}$}=
1.37$\pm$0.13}) and $\gamma$-ray ({ High State: $\rm\alpha$ = -2.29$\pm$0.01,
$\rm\beta$ = 0.086$\pm$0.004; Low State: $\rm\alpha$ = -2.40$\pm$0.01, 
$\rm\beta$ = 0.30$\pm$0.02)} points are shown at the 
index-weighted\footnote{$\rm <E> = \int E N(E)dE/\int N(E)dE$}
mean energy, where the indices are the mean of respective flux states.

\begin{table}[!ht]
 \centering
 \caption{SED parameters for Low State and High State}
 \begin{tabular}{l c c} 
 \hline
  Parameters & \multicolumn{2}{c}{Numerical values (CGS units)}	 \\ 
	     & Low State 		& High State \\ \hline
  Particle index before break (p)	& 2.0	& 1.8 \\
  Particle index after break (q)	& 4.1	& 4.0 \\
  Magnetic field ($\rm B^\prime$, G)	& 0.77	& 0.88 \\
  Equipartition factor ($\rm \eta$)	& 3.4	& 1.9 \\
  Doppler factor ($\rm \delta$)		& 19.5	& 30.1 \\
  Particle break energy$^\ast$ ($\rm \gamma_{\rm b^\prime}$)	& $8.7 \times 10^2$ & $5.5 \times 10^2$	\\
  Jet power ($P_{\rm jet}$)		& $ \rm 4.3\times 10^{46}$ & $ \rm 3.8\times 10^{46}$	\\
 \hline
 Emission region size ($\rm R^\prime$): $1 \times 10^{16}$ cm 	\\
 Angle to the line of sight ($\rm theta$): 3$^\circ$ \\
 IR Torus temperature ($\rm T_\ast$): 820 K \\
 Minimum particle energy$^\ast$ ($\gamma_{\rm min}^\prime$): 10 \\
 Maximum particle energy$^\ast$ ($\gamma_{\rm max}^\prime$):  $5 \times 10^4$	\\
 $^\ast$in electron rest-mass units
 \end{tabular}
 \label{tab:SEDparVal}
\end{table}

\section{DISCUSSION}\label{sec:discussion}
Our systematic analysis of multi-wavelength flux distribution of FSRQ
\source, having well sampled flux states (Figure \ref{fig:histComSpec}
top panel) reveals that $\gamma$-rays to IR-optical emission follow
a bi-lognormal distribution except X-ray\footnote{A hint of bi-lognormality
is present in \emph{Swift}-XRT data but the data are biased.},
which conforms to a single lognormal distribution. Such bimodality at
$\gamma$-ray energies has been expected from extragalactic $\gamma$-ray background
(EGB) study \citep[see also \citet{2011ApJ...736...40S}]{1996ApJ...464..600S}.
Interestingly, the amplitude ratio is $\sim 4$, close to the one anticipated
($\sim 5$) by \citet{1996ApJ...464..600S}. Further, the mean spectral index
during the High State (2.29$\pm$0.01, 0.086$\pm$0.004) is relatively harder than
the Low State (2.40$\pm$0.01, 0.30$\pm$0.02), but both ($\alpha$ only) are consistent with the
observed distribution of indices inferred from the statistical studies of LAT blazars
\citep{2012ApJ...753...45S, 2015MNRAS.454..115S}. In terms of contribution to EGB,
blazars account for almost all above 100 GeV and $\sim$ 50\% (FSRQ: $\sim$ 35\%,
BL-Lac: $\sim$ 17\%) below it with the rest accounted by other LAT detected sources
\citep{2015ApJ...800L..27A, 2015MNRAS.454..115S}.

Similar studies, performed so far on BL Lac class of sources have shown both
single as well as bi-lognormal distributions. The X-ray profile found here is also
seen in BL Lacertae \citep{2009A&A...503..797G}. Similarly, a single lognormal profile
in multi-wavelength have been found for PKS 2155-304 \citep{2015arXiv150903104C} except
for the single flaring instance of 2006 at $>200$ GeV, inclusion of which results in a bi-lognormal
profile at $\gamma$-rays \citep{2010A&A...520A..83H}. Here, however, we have found flux distribution
with a bi-lognormal distribution at IR-optical as well as $\gamma$-rays (0.1-300 GeV) with
light curves showing many flaring instances of varied flux levels in different energy bands. 
This bi-modality seems to be similar to the two basic spectral states, (low/hard and
high/soft) that are observed in X-ray binaries \citep[e.g.][]{2005MNRAS.359..345U}. However,
these spectral transitions in X-ray binaries occur on timescales of days/months. Scaling this time-scale for
a $10^7$ SMBH would lead to $10^{4-5}$ years \citep{2006MNRAS.372.1366K}, significantly longer
than the days-to-weeks transitions reported here.

The intrinsic rms variability during the two states exhibits a complex energy
dependent behavior (Figure \ref{fig:histComSpec}, bottom-left) with remarkably
different multi-wavelength flux dispersions. The $\gamma$-ray Low and High State
flux dispersions are consistent with each other (Table \ref{tab:2GaussPar}). The same
is true for the B-band dispersions. However, the $\gamma$-ray dispersion is comparatively
wider than the IR-optical and X-ray during the Low State, but similar to IR-optical
dispersions during the High State except for the B-band. Further, the dispersion
increases towards IR bands during the Low State
suggesting more variability at IR bands than optical. Additionally, the IR-optical
distributions exhibit significant high flux tails (Fig. \ref{fig:histComSpec},
top-right panel), even after averaging over LAT binning durations, contrary to the
X-ray and $\gamma$-ray distributions. 

The SED modeling requires both BLR and IR-torus \citep[$\sim$ 820 K,][]{2008ApJ...672..787K}
to reproduce the 0.1-10 GeV emission in leptonic scenario with bulk Lorentz factor
being mainly responsible for variation. This is also consistent with $\gamma$-ray
dispersion being wider (BLR-IR boosted by square of bulk Lorentz factor) than
IR-optical which is synchrotron radiation. On the other hand, similar High State
dispersions at all energies except B-band suggest both underlying particle spectrum and
magnetic field playing a dynamic role. The high flux tails of IR-optical
distributions, however, favor magnetic field, indicating a shock origin for the high
flux states. Similar High and Low State dispersions in B-band, on the other hand,
is result of disk emission being dominant during both as suggested
by the High State SED and RMS spectrum (Fig. \ref{fig:histComSpec}).
The single lognormal distribution for X-ray from RXTE is consistent with 3-20 keV
flux not as strongly variable as the IR-optical and $\gamma$-ray emission and is mainly
a result of emission from low energy part of the particle spectrum. We disfavor
a substantial coronal contribution at X-ray for single lognormal distribution as even in
that case a bimodal distribution is expected as seen in the optical B-band (corona
connected with the disk).

A lognormal distribution of flux implies emission being powered by cascade/multiplicative
processes rather than additive. In black hole powered sources like Seyfert galaxies
and X-ray binaries, lognormality is widely believed to be a result of fluctuations
in the accretion disk \citep{2005MNRAS.359..345U, 2008bves.confE..14M}. Thus, a
lognormal flux distribution in blazars may be an imprint of modulation in the disk,
connecting jet to disk activity \citep{2009A&A...503..797G, 2008bves.confE..14M}.
Additionally, a lognormal distribution is also expected for plasma ejection
triggered by local magnetic reconnections  \citep[see eg.][]{2003ICRC....5.2729A},
but contribute negligibly to blazars emission \citep{2015ApJ...802..113K}.
Most of emission is from the jet \citep{2015ApJ...802..113K}, mainly from the 
blazar zone where the plasma collimates, probably by the action of large-scale
guiding magnetic fields and/or external medium, as inferred from the M87
jet \citep{2013ApJ...775...70H}. The cascades of magnetic energy to particles
proceed probably via reconnection first, leading to further development
via turbulence and shocks. A reconnection mediated channeling of energy to particles,
however, results at most to an equipartition of energy between the two, in contradiction
with the SED modeling of blazars emission which almost always suggests particle
dominated jets \citep{2014Natur.515..376G}. It is, though, consistent with
relativistic fluid approach of \citet{2015MNRAS.453.4070P}
(see also \citet{2014ApJ...782...82D}).

A lognormal distribution also suggests an existence of some characteristic energy scale
that manifests into radiative form. This may be related to the power being injected into
the jet. The apparent large flux variations, on the other hand, may be related to the efficiency
of underlying processes channeling energy among themselves and finally to non-thermal
particles at the blazar zone and/or within the jet. Apart from active BHs, coronal mass
ejection from the Sun, known to be a result
of magnetic reconnection also follow a lognormal distribution \citep{2003ICRC....5.2729A}.
The observed lognormality, thus, may well be an imprint of disk modulation as seen
in X-ray binaries and Seyfert galaxies, and/or a convolution of plasma exhaust due to
local magnetic reconnections.

The authors thank the referee for a thorough report. This research has made use of
data obtained from High Energy Astrophysics Science
Archive Research Center (HEASARC), maintained by NASA’s Goddard Space Flight Center
and an up-to-date SMARTS optical/near-infrared light curves available at
www.astro.yale.edu/smarts/glast/home.php.

\end{document}